\documentclass[aps,prx,amsmath,superscriptaddress,twocolumn,longbibliography]{revtex4-1}

\usepackage{graphicx}
\usepackage{amsmath}
\usepackage{amsfonts}
\usepackage{amssymb}
\usepackage{braket}
\usepackage{bm}
\usepackage{multirow}
\usepackage{color}
\usepackage[normalem]{ulem}
\usepackage{mathrsfs}
\usepackage{mathtools}
\usepackage{float}

\makeatletter

\newcommand{\Rmnum}[1]{\expandafter\@slowromancap\romannumeral #1@}
\makeatother

\usepackage[breaklinks]{hyperref}
\hypersetup{colorlinks=true, linkcolor=blue, citecolor=blue, filecolor=blue, urlcolor=blue}

\AtBeginDocument{%
    \newwrite\bibnotes
    \def\bibnotesext{Notes.bib}
    \immediate\openout\bibnotes=\jobname\bibnotesext
    \immediate\write\bibnotes{@CONTROL{REVTEX41Control}}
    \immediate\write\bibnotes{@CONTROL{%
    apsrev41Control,author="08",editor="1",pages="1",title="0",year="1"}}
     \if@filesw
     \immediate\write\@auxout{\string\citation{apsrev41Control}}%
    \fi
}%

\begin{document}

\title{Extended nonergodic regime and spin subdiffusion in disordered SU(2)-symmetric Floquet systems}

\author{Zhi-Cheng Yang}
\email{zcyang@umd.edu}
\affiliation{Joint Quantum Institute, University of Maryland, College Park, MD 20742, USA}
\affiliation{Joint Center for Quantum Information and Computer Science, University of Maryland, College Park, MD 20742, USA}
\author{Stuart Nicholls}
\author{Meng Cheng}
\email{m.cheng@yale.edu}
\affiliation{Department of Physics, Yale University, New Haven, CT 06520-8120, USA}

\date{\today}

\begin{abstract}

We explore thermalization and quantum dynamics in a one-dimensional disordered SU(2)-symmetric Floquet model, where a many-body localized phase is prohibited by the non-abelian symmetry. Despite the absence of localization, we find an extended nonergodic regime at strong disorder where the system exhibits nonthermal behaviors.
In the strong disorder regime, the level spacing statistics exhibit neither a Wigner-Dyson nor a Poisson distribution, and the spectral form factor does not show a linear-in-time growth at early times characteristic of random matrix theory. The average entanglement entropy of the Floquet eigenstates is subthermal, although violating an area-law scaling with system sizes. We further compute the expectation value of local observables and find strong deviations from the eigenstate thermalization hypothesis. The infinite temperature spin autocorrelation function decays at long times as $t^{-\beta}$ with $\beta < 0.5$, indicating subdiffusive transport at strong disorders.

\end{abstract}

\maketitle

\section{Introduction}

The presence of strong quenched disorders in non-integrable quantum systems often impedes thermalization. A prototypical example in one dimension is many-body localization (MBL)~\cite{PhysRevB.75.155111, PhysRevB.82.174411, nandkishore2015many}, where highly-excited eigenstates of the Hamiltonian violate the eigenstate thermalization hypothesis (ETH)~\cite{PhysRevA.43.2046, PhysRevE.50.888} and the system fails to reach thermal equilibrium starting from generic initial states having a finite energy density. However, disorders have a dramatic effect on the quantum dynamics even in the ergodic phase of systems exhibiting a MBL phase transition. It has been shown that thermalization in this regime is anomalous, featuring subdiffusive transport as well as slow relaxation of local observables to their thermal expectation values~\cite{PhysRevLett.114.100601, PhysRevX.5.031033, PhysRevB.93.060201, PhysRevB.93.224205, PhysRevB.94.045126, PhysRevB.89.220201, PhysRevLett.114.160401, PhysRevLett.117.040601, luitz2017ergodic, PhysRevLett.117.170404, PhysRevB.96.104201, PhysRevB.98.180201}.

On the other hand, a true MBL phase is incompatible with non-abelian global symmetries, such as an SU(2) spin rotation symmetry~\cite{PhysRevB.96.041122, PhysRevB.94.224206, PhysRevLett.114.217201, PhysRevX.10.011025, PhysRevB.102.035117, glorioso2020hydrodynamics}. The quasi-local integrals of motion, a defining feature of MBL, form exactly degenerate multiplets under a non-abelian symmetry group, which are unstable against any infinitesimal interactions between them, and hence must break down.
One may thus naively expect that systems with non-abelian symmetries are trivially thermal even at strong disorders. However, Ref.~\cite{PhysRevX.10.011025} recently studied an SU(2)-symmetric random Heisenberg chain using a real-space renormalization group approach and identified a broad regime in system sizes where the system appears nonergodic. Within this regime, the eigenstates are well-approximated by tree tensor networks with faster than area-law but strongly subthermal entanglement entropy scaling, and expectation values of local observables exhibit deviations from generic thermalizing systems. The extremely long length scale beyond which the eventual thermalization sets in makes this intermediate nonergodic regime directly relevant in typical experiments with moderate system sizes.

While the approach taken in Ref.~\cite{PhysRevX.10.011025} depends crucially on energetics in a Hamiltonian system, it is natural to ask what will happen in a strongly disordered periodically-driven Floquet system with SU(2) symmetry, where energy conservation is absent and a tree tensor network structure for the eigenstates does not seem to hold. The lack of energy conservation tends to allow Floquet systems to thermalize faster and more completely to infinite temperature than Hamiltonian systems~\cite{PhysRevX.4.041048, ponte2015periodically, PhysRevB.94.224202, russomanno2015thermalization, PhysRevE.90.012110, PhysRevA.92.062108}. Previous studies suggest that thermalization and transport in Floquet systems near a MBL transition are anomalous~\cite{PhysRevB.94.094201, PhysRevB.93.134206, PhysRevB.98.060201, PhysRevE.97.022202, notarnicola2020slow}. However, it is unclear whether such extended nonergodic regime becomes more fragile once the nearby MBL phase is absent.
Moreover, the transport property in disordered systems with SU(2) symmetry remains an open question.

In this work, we address the above questions by studying a one-dimensional disordered SU(2)-symmetric Floquet model. The key properties of a time periodic Hamiltonian $H(t+T) = H(t)$ are encoded in the eigenstates of the Floquet operator $U_F = \mathcal{T} e^{-i \int_0^T dt H(t)}$, which generates time evolution over integer multiples of periods. We first look at the level spacing statistics of the eigenenergy spectrum of $U_F$ using exact diagonalization and find no transition into a MBL phase, which is consistent with the SU(2) symmetry. However, at strong disorder, the level spacing statistics exhibit neither a Wigner-Dyson nor a Poisson distribution, and the drift towards a Wigner-Dyson distribution upon increasing the system size is very slow. To further probe the long-range spectral correlations beyond nearest-neighboring levels, we calculate the disorder averaged spectral form factor. We find that the spectral form factor also deviates from random matrix behaviors. In particular, within system sizes accessible numerically, the linear-in-time growth at early times is absent, and the curves coincide with random matrix theory predictions only at timescales comparable to the inverse level spacings. The average entanglement entropy of the Floquet eigenstates is subthermal, although exhibiting a faster than area-law scaling with system sizes.

The spectral properties of the Floquet model suggest an intermediate regime that is neither MBL nor quantum chaotic in the usual sense. To directly test ETH in this regime, we calculate the distribution of the expectation values of local observables under Floquet eigenstates, and find that it strongly deviates from the Floquet version of ETH, which predicts a Gaussian distribution centered around the infinite temperature average value. This suggests the existence of an extended delocalized and yet nonergodic regime~\cite{ PhysRevB.101.064302, russomanno2020non}.
Finally, we study transport properties in the strong disorder regime by computing the infinite temperature spin autocorrelation function, which decays at long times as $t^{-\beta}$ with $\beta < 0.5$, indicating that spin transport is subdiffusive at strong disorders.

\section{The model}

We consider a spin-1/2 system with Heisenberg interactions that respect the global SU(2) spin rotation symmetry. Time evolutions are generated by switching between two alternating Hamiltonians:
\begin{eqnarray}
H_1 &=& \sum_i \sqrt{1-g^2} J_i \ {\bm S}_i \cdot {\bm S}_{i+1}  \nonumber  \\
H_2 &=& \sum_i  g \ J_2 \ {\bm S}_i \cdot {\bm S}_{i+2},
\label{eq:model}
\end{eqnarray}
with the Floquet operator given by:
\begin{equation}
U_F = {\rm exp} \left(-i H_2 T/2\right) \ {\rm exp} \left( -i H_1 T/2 \right)
\end{equation}
with period $T$. $J_i$'s are Gaussian distributed random nearest-neighbor couplings with zero mean and variance unity, and $J_2$ is uniform across all next-nearest-neighboring spins. We included the $J_2$ term to make the model more generic while respecting the symmetry. The parameter $0<g<1$ controls the relative strength between the disordered nearest-neighbor couplings and the uniform next-nearest-neighbor coupling, thereby effectively tuning the disorder strength. We choose the above normalization such that the many-body bandwidth remains fixed as $g$ varies. The limit when $g=1$ is integrable, corresponding to two copies of the clean Heisenberg model on odd and even sites respectively. However, taking $g$ to be strictly less than one makes the model nonintegrable. A disorder-free Hamiltonian system corresponding to $H_1+H_2$ has been shown to thermalize, with the equilibrium distribution described by a non-abelian thermal ensemble~\cite{PhysRevE.101.042117}.
We hereafter choose $J_2=1$ and $T=4$ for numerical simulations. With this choice of parameters, the principal quasienergy zone width $\frac{2\pi}{T}$ is smaller than the bandwidth of the undriven Hamiltonians $H_1$ and $H_2$. Hence, we are exploring the low driving frequency regime where the Floquet Hamiltonian behaves differently than the undriven model~\cite{PhysRevLett.115.030402}.

As a result of the SU(2) symmetry, model~(\ref{eq:model}) has two commuting conserved quantities: $({\bm S}_{\rm tot}^2, S_{\rm tot}^z)$ with ${\bm S}_{\rm tot} = \sum_i {\bm S}_i$ being the total spin and $S_{\rm tot}^z$ being its $z$-projection. Therefore, the Hilbert space of our model falls into distinct blocks labeled by two quantum numbers $S_{\rm tot}^z$ and ${\bm S}_{\rm tot}^2 = S(S+1)$. For simplicity, we shall focus on the sector with $S_{\rm tot}^z=0$ and total spin $S=0$.

\section{Spectral statistics}

We shall now present our numerical results on the spectral statistics of the quasi-energy spectrum of the Floquet operator $U_F$, and demonstrate that the spectrum at strong disorder shows deviations from random matrix theory predictions.

\subsection{Level spacing statistics}

The analog of eigenenergies in a Floquet system is given by the eigenvalues of the Floquet operator, which are unimodular and can be denoted as $\{ e^{i \theta_n} \}$. The quasi-energies $\{\theta_n \}$ are $2\pi$ periodic, hence we take them to be within the principal zone $[-\pi, \pi)$. Let $\{ \theta_n \}$ be rank-ordered descendingly, such that $\theta_n > \theta_{n+1}$, and define the gap between adjacent quasi-energy levels as $\Delta \theta_n = \theta_{n-1}-\theta_n >0$. The level spacing distribution can be captured by the ratio between adjacent gaps:
\begin{equation}
r_n = \frac{{\rm min}(\Delta \theta_n, \Delta \theta_{n+1})}{{\rm max}(\Delta \theta_n, \Delta \theta_{n+1})}.
\end{equation}
The average value of $r_n$ over different levels is able to capture the distributions of level spacings, and since Floquet quasi-energies have a uniform spectral density, we take the average over the entire spectrum.
This quantity serves as the canonical diagnostic for the phase transition between a thermalizing phase and MBL phase~\cite{PhysRevB.75.155111, PhysRevB.82.174411}. In the localized phase with a Poisson distributed spectrum, $\langle r \rangle \approx 0.39$; in the thermalized phase, the quasi-energy spectrum of our model follows a circular orthogonal ensemble (COE) with $\langle r \rangle \approx 0.53$, since both $H_1$ and $H_2$ are time reversal symmetric~\cite{PhysRevX.4.041048, PhysRevB.93.104203}.

\begin{figure}[!t]
\includegraphics[width=.45\textwidth]{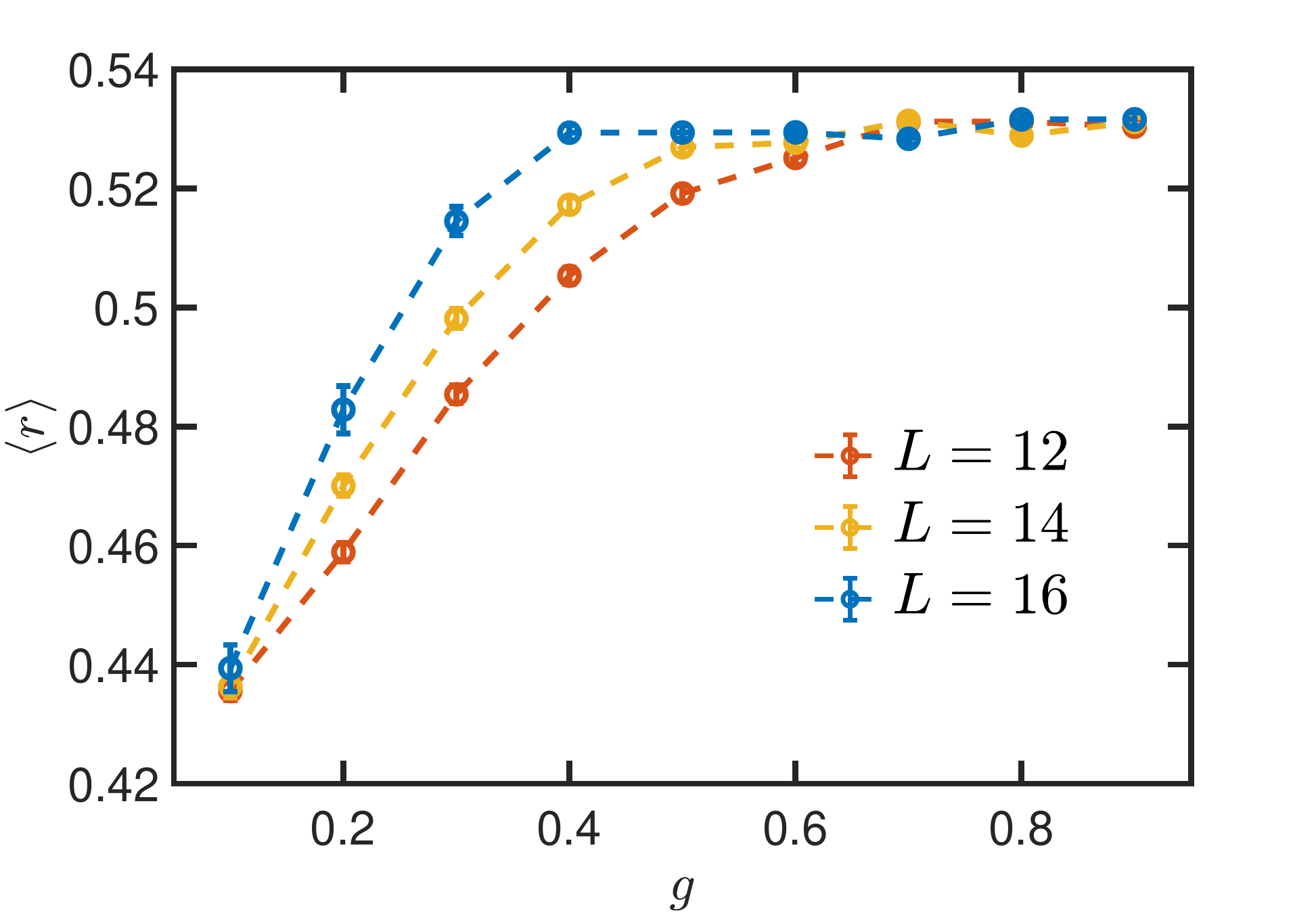}
\caption{The average level spacing ratio as a function of the disorder strength $g$ for different system sizes. Each data point is averaged over 500 disorder realizations for $L=12$, 250 realizations for $L=14$, and $50$ realizations for $L=16$.}
\label{fig:levelspacing} 
\end{figure}

In Fig.~\ref{fig:levelspacing}, we show the average level spacing ratio $\langle r \rangle$ as a function of the disorder strength $g$ for different system sizes. First of all, we find that there is no transition into a MBL phase, as expected for SU(2)-symmetric systems in general. Second, the flow of $\langle r \rangle$ upon increasing system sizes is always monotonic, as opposed to the Hamiltonian system studied in Ref.~\cite{PhysRevX.10.011025}. This is due to the removal of energy conservation as well as the uniform spectral density in Floquet systems, which allows one to take all quasi-energies into the average on equal footing. At weak disorders, $\langle r \rangle$ approaches the COE value, indicating that the system is nonintegrable and thermalizing. On the other hand, at strong disorder, $\langle r \rangle$ approaches a value that is intermediate between a Poisson and COE distribution, with short-range level repulsion. While one expects that the system will eventually thermalize in the thermodynamic limit, for system sizes accessible in our numerics, the flow towards COE with increasing system sizes is extremely slow. It is therefore possible that there is an extended nonergodic regime at strong disorders, similar to the strong disorder regime on the ergodic side of a MBL transition~\cite{PhysRevB.89.220201, PhysRevLett.114.160401, PhysRevLett.117.040601, luitz2017ergodic, PhysRevLett.117.170404, PhysRevB.96.104201, PhysRevB.98.180201}.

\begin{figure}[!t]
\centering
\includegraphics[width=.47\textwidth]{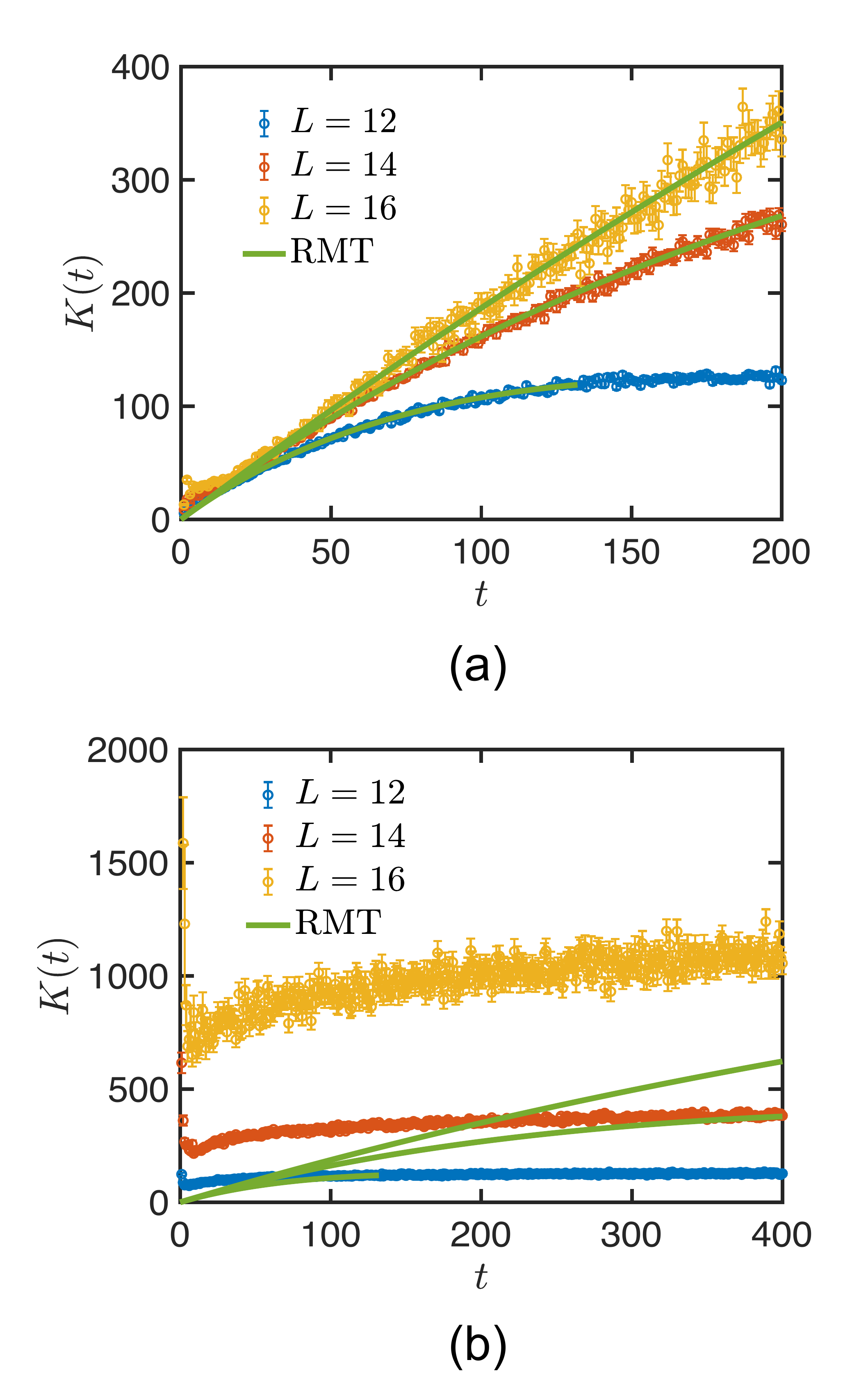}
\caption{The spectral form factor $K(t)$ for (a) $g=0.9$ (weak disorder) and (b) $g=0.1$ (strong disorder). The green curves indicate the random matrix theory prediction Eq.~(\ref{eq:Kcoe}) for COE. The results are averaged over 2000 realizations for $L=12, 14$, and 500 realizations for $L=16$.}
\label{fig:formfactor} 
\end{figure}

\subsection{Spectral form factor}

While the level spacing statistics capture the repulsion between nearest-neighboring quasi-energy levels, we now consider a complementary spectral measure that is capable of describing correlations beyond nearest-neighbor levels. The spectral form factor of the quasi-energy spectrum is defined as:
\begin{equation}
K(t) = \left\langle \sum_{i,j}  e^{i (\theta_i - \theta_j)t} \right\rangle,
\label{eq:formfactor}
\end{equation}
where the average is taken over different disorder realizations. This quantity is intimately related to the temporal two-point correlation functions of local observables, and has been playing a central role in characterizing quantum chaos~\cite{cotler2017chaos, PhysRevLett.121.060601, PhysRevD.98.086026}. Since the definition~(\ref{eq:formfactor}) involves all pairs of quasi-energy levels, it is able to capture spectral correlations beyond the scale of level spacing. For orthogonal ensembles, $K(t)$ takes the form~\cite{mehta2004random}:
\begin{equation}
K(t) =
\begin{cases}
\mathcal{N} \left[2 \tau - \tau {\rm ln}(1+2 \tau )\right]   &   (\tau \leq 1); \\
\\
\mathcal{N} \left[ 2 - \tau {\rm ln} \left( \frac{2\tau+1}{2\tau-1}\right)\right]  &  (\tau>1),
\end{cases}
\label{eq:Kcoe}
\end{equation}
where $\mathcal{N}$ is the Hilbert-space dimension, and $\tau = t/\mathcal{N}$. Notice that the behavior of $K(t)$ in Eq.~(\ref{eq:Kcoe}) is different from that in the unitary ensembles, where $K(t)$ is simply a linear ramp for $\tau<1$ followed by a plateau for $\tau>1$~\cite{mehta2004random, cotler2017chaos, PhysRevLett.121.060601, PhysRevD.98.086026}.
At short times, expanding Eq.~(\ref{eq:Kcoe}) at small $\tau$ yields $K(t) \approx 2t$. Thus at early times, $K(t)$ grows linearly in time with a different slope from the unitary ensembles.

In Fig.~\ref{fig:formfactor}, we plot the spectral form factor of our model at weak and strong disorders. One can see that at weak disorders, the spectral form factor agrees very well with random matrix theory predictions. On the other hand, at strong disorders, $K(t)$ strongly deviates from Eq.~(\ref{eq:Kcoe}). In particular, the linear-in-$t$ growth at early times is absent, and the curves only agree with the random matrix theory behavior at late times comparable to the Heisenberg timescale $\sim \mathcal{N}$. This implies that the long range spectral correlations in the quasi-energy spectrum does not follow the random matrix theory behavior. For the system sizes accessible in our numerics, level repulsion between quasi-energy levels exists only within the order of a few level spacings, as indicated by the non-Poissonian level spacing ratio.

\begin{figure}[!t]
\includegraphics[width=.47\textwidth]{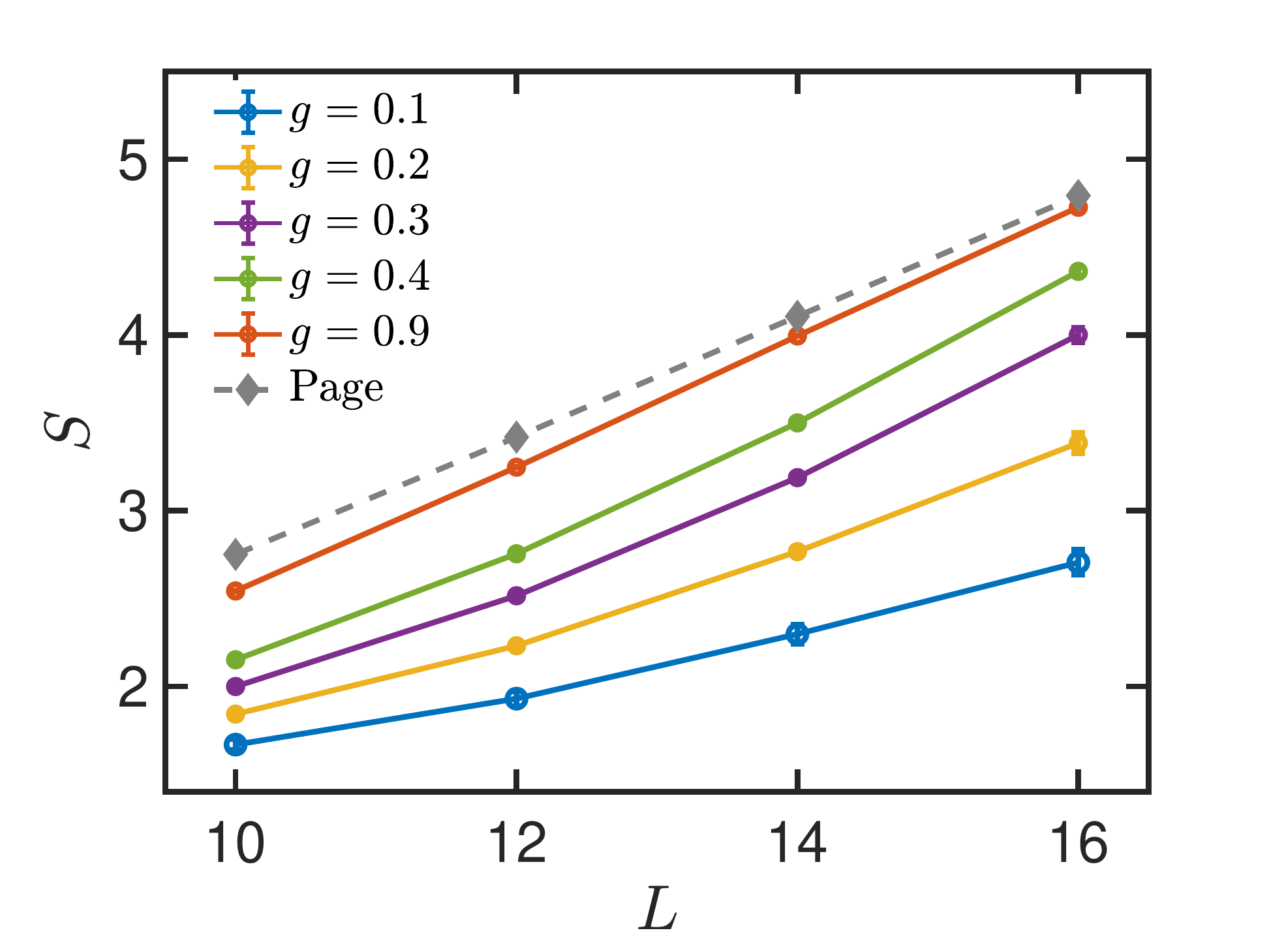}
\caption{Average entanglement entropy as a function of system sizes, for different disorder strengths. The disorder is strong for small $g$ and weak for large $g$. At strong disorders, the entanglement entropy scales faster than area-law, but has a subthermal value. The Page values corresponding to random states within the $S^z_{\rm tot}=0$ and $S=0$ sector are shown in dashed line.
The numbers of realizations in obtaining each data point are the same as in Fig.~\ref{fig:levelspacing}.}
\label{fig:entropy} 
\end{figure}

\section{Entanglement entropy and local observables}

We next turn to the entanglement entropy scaling and local observables of our model. In Fig.~\ref{fig:entropy}, we plot the entanglement entropy averaged over all eigenstates as a function of system sizes. We take an equi-bipartitioning of the system in the middle, and compute the von-Neumann entropy:
\begin{equation}
S = - {\rm Tr} \left( \rho_A {\rm ln} \rho_A\right),
\end{equation}
where $\rho_A = {\rm Tr}_{\bar{A}} |\psi\rangle \langle \psi|$ is the reduced density matrix of subsystem $A$. We find that the entanglement entropy at strong disorders scales faster than area-law, which is a constant in one dimension. This is again consistent with the general expectation that the system is not many-body localized at strong disorders. However, the values of the entanglement entropy for small $g$ are well below the (infinite temperature) thermal values for the given size of Hilbert space. This indicates that, although the system does not localize at strong disorders, it also does not heat up to infinite temperature, as would be the case for generic thermalizing Floquet systems.
This implies that although the eigenstates are delocalized in this regime, they are nevertheless nonergodic. On the other hand, at weak disorders the entanglement entropy is close to the Page value~\cite{PhysRevLett.71.1291} corresponding to a random state within the $S^z_{\rm tot}=0$ and $S=0$ sector.

To further characterize the nonergodic regime at strong disorders, we study expectation values of local observables as a direct test for ETH. For Hamiltonian systems, ETH suggests that the diagonal matrix element of local observables is a continuous function of the energy, and centered around its microcanonical ensemble average value. The Floquet version of ETH thus implies that the expectation value of local observables under Floquet eigenstates should be narrowly peaked around their infinite temperature average value, with fluctuations decaying exponentially with increasing system sizes. We choose the observable ${\bm S}_i \cdot {\bm S}_{i+1}$ associated with a particular bond between site $i$ and $i+1$. The eigenvalue of ${\bm S}_i \cdot {\bm S}_{i+1}$ equals to $-\frac{3}{4}$ for singlet bonds, and $\frac{1}{4}$ for triplet bonds.

\begin{figure*}[!t]
\includegraphics[width=.9\textwidth]{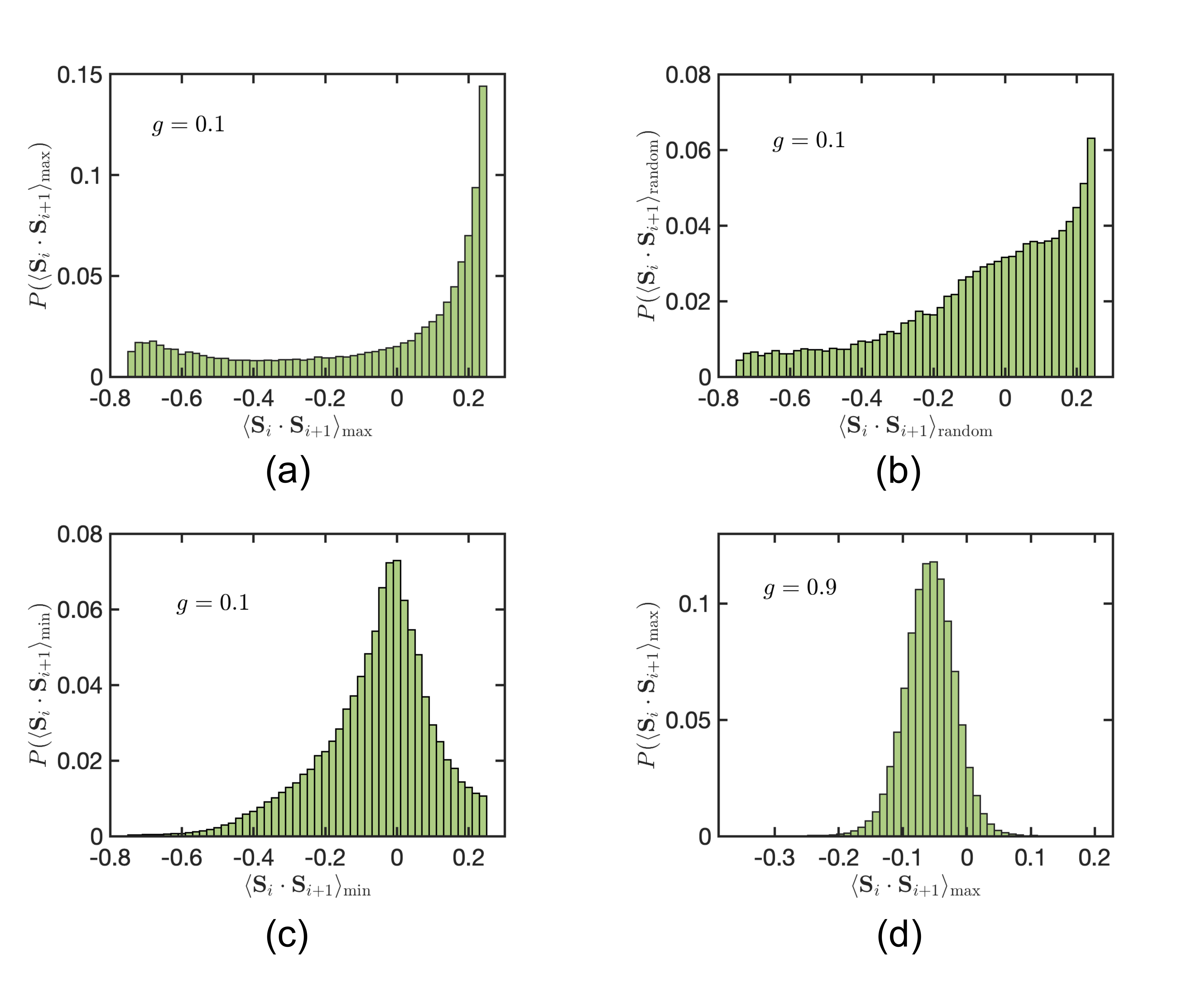}
\caption{Probability distributions of the local observable ${\bm S}_i \cdot {\bm S}_{i+1}$ over all eigenstates for strong disorders $g=0.1$ [(a), (b) \& (c)] and weak disorders $g=0.9$ [(d)]. In (a) \& (d), the strongest bonds with the largest $|J_i|$ are picked; in (c), the weakest bonds with the smallest $|J_i|$ are picked; in (b) the bonds are picked randomly. The distributions are obtained for system size $L=14$ and over 250 disorder realizations.}
\label{fig:bond} 
\end{figure*}

In Fig.~\ref{fig:bond}(a)-(c), we plot the probability distributions of $\langle {\bm S}_i \cdot {\bm S}_{i+1} \rangle$ at strong disorders across all eigenstates for different choices of bonds. In Fig.~\ref{fig:bond}(a), we choose the strongest bonds with the largest $|J_i|$ for each disorder realization. We find that the distribution is nearly bimodal, with dominating weights centered around $\frac{1}{4}$ and a weaker peak around $-\frac{3}{4}$. This clearly shows that the pair of spins coupled via the strongest bond almost form a triplet or singlet instead of thermalizing with the rest of the system. One expects that the probability of finding a triplet or a singlet bond is proportional to the number of multiplets $(2S+1)$, which is apparently bigger for triplets with $S=1$. This explains the peak around $\frac{1}{4}$ in Fig.~\ref{fig:bond}(a), which corresponds to triplet bonds. In contrast, for weak disorders, the probability distribution for even the strongest bond is a Gaussian centered around the infinite temperature average value with a narrow width, as shown in Fig.~\ref{fig:bond}(d), in agreement with ETH.

\begin{figure}[t]
\includegraphics[width=.43\textwidth]{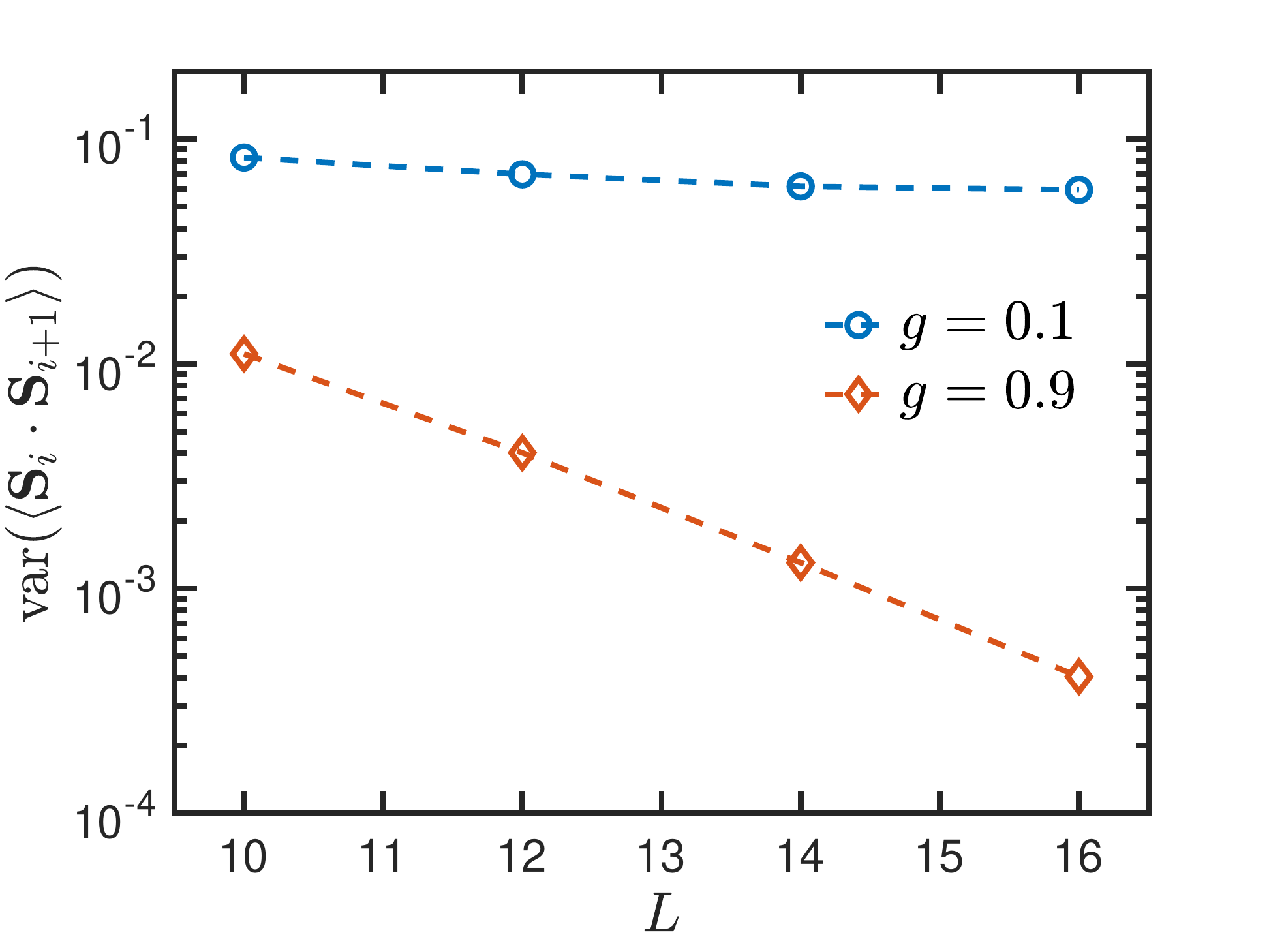}
\caption{Variance of the expectation values of ${\bm S}_i \cdot {\bm S}_{i+1}$ on randomly chosen bonds for strong and weak disorders. The ensemble includes both different Floquet eigenstates and disorder realizations.}
\label{fig:fluctuation} 
\end{figure}

We further show the probability distributions for the weakest bond with the smallest $|J_i|$ [Fig.~\ref{fig:bond}(c)], and a randomly chosen bond [Fig.~\ref{fig:bond}(b)]. In Fig.~\ref{fig:bond}(b), we find that for a randomly chosen bond, the probability distribution also deviates from ETH behavior, featuring a broad non-Gaussian distribution within its domain. To quantify the ultimate approach to ETH in the thermodynamic limit, we compute the variance of local observable ${\bm S}_i\cdot {\bm S}_{i+1}$ for a randomly chosen bond as a function of system sizes, as shown in Fig.~\ref{fig:fluctuation}. For weak disorders $g=0.9$, the fluctuations decay exponentially with increasing system sizes, consistent with ETH. On the other hand, the fluctuations at strong disorders are significantly larger than the weak disordered case. Our numerics indicate that the fluctuations at strong disorders also decay as the system size increases, although much more slowly than the weak disordered case. Due to this extremely slow decay, we cannot tell from numerics on small system sizes whether or not the asymptotic decay is exponential in system size.
Finally, for the weakest bond for each disorder realization, the expectation value is no longer peaked around that of a triplet. Instead, it is now centered near zero, with only small weights around the singlet and triplet values. This implies that pairs of spins that are weakly coupled do not form singlets or triplets between themselves and tend to thermalize with the rest of the system. However, comparing with Fig.~\ref{fig:bond}(d), the distribution still shows deviations from ETH predictions, namely, the distribution in Fig.~\ref{fig:bond}(c) has heavy tails away from its peak. 

Therefore, we conclude that the diagonal matrix element of ${\bm S}_i \cdot {\bm S}_{i+1}$ exhibits deviations from ETH predictions. Its expectation value shows a broad non-Gaussian distribution, with fluctuations decaying much more slowly than predicted by ETH. Spins that are strongly coupled tend to form triplets, and even in the presence of driving, such strongly coupled pairs have a hard time absorbing energy from the drive and hence are nearly decoupled from the rest of the system. Expectation values associated with a typical bond also show a broad distribution within their domain.

\section{Spin autocorrelation function}

In this section, we study the transport property of our model. Since the total $z$-magnetization is conserved, one can thus focus on transport of local magnetizations. For disordered non-integrable systems that are thermalizing, one usually expects that the transport of the conserved charge should be diffusive. However, several studies have found anomalous subdiffusion behavior in the ergodic regime of systems exhibiting a MBL phase transition~\cite{PhysRevLett.114.100601, PhysRevX.5.031033, PhysRevB.93.060201, PhysRevB.93.224205, PhysRevB.94.045126, PhysRevB.89.220201, PhysRevLett.114.160401, PhysRevLett.117.040601, luitz2017ergodic, PhysRevLett.117.170404, PhysRevB.96.104201, PhysRevB.98.180201, PhysRevB.94.094201, PhysRevB.93.134206, PhysRevB.98.060201}. Here we present numerical evidence of a similar subdiffusive regime in our model, despite the absence of a true MBL phase.

\begin{figure}[t]
\includegraphics[width=.45\textwidth]{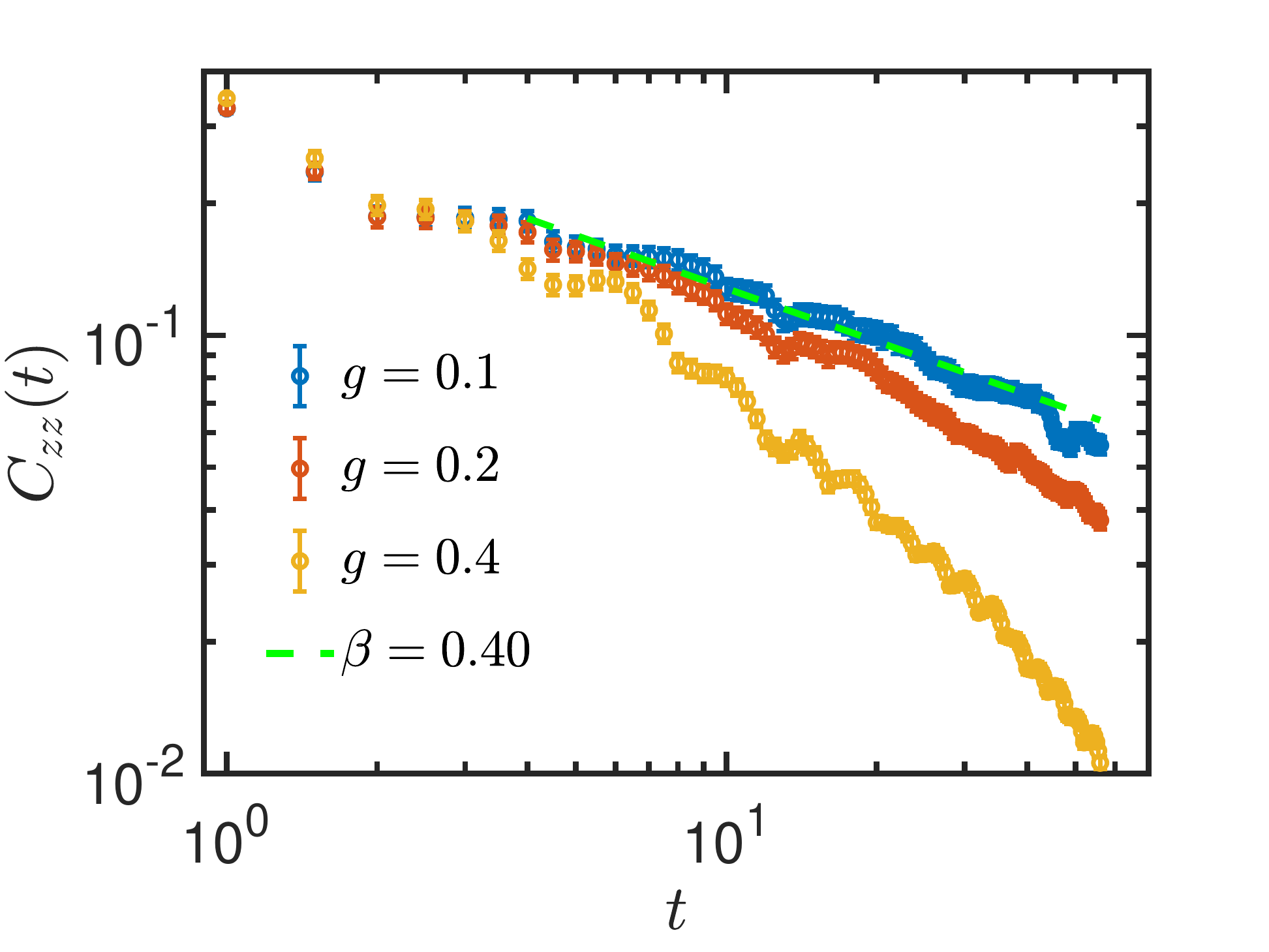}
\caption{Spin autocorrelation function $C_{zz}(t)$ for different disorder strengths with $L=20$ and averaged over 500 disorder realizations. For strong disorders ($g=0.1$), spin transport is subdiffusive with $C_{zz}(t) \sim t^{-0.4}$. For weak disorders ($g=0.4$), one appears to find $\beta > 0.5$ suggesting superdiffusion, but this is due to finite-size effect, as explained in the text.}
\label{fig:correlation} 
\end{figure}

We consider the infinite temperature spin autocorrelation function:
\begin{equation}
C_{zz}(t) = \frac{1}{\mathcal{N}} {\rm Tr} \left[ S^z_i(t) S^z_i(0)\right],
\end{equation}
where the trace is taken within a fixed total magnetization sector. Physically, this autocorrelation function probes the probability of finding an initially localized charge at the same position at time $t$. At late times, $C_{zz}(t)$ decays as a power law: $C_{zz}(t) \sim t^{-\beta}$, where $\beta=0.5$ for diffusion and $0<\beta<0.5$ for subdiffusion. We compute the above autocorrelation function using standard Krylov space time evolution methods~\cite{luitz2017ergodic}, and the results are shown in Fig.~\ref{fig:correlation}. We find that the spin transport at strong disorder has $\beta \approx 0.4$ and hence is subdiffusive. Notice that the fitting of the power $\beta$ from $C_{zz}(t)$ is typically not extremely accurate, due to the oscillations on top of the power-law decay as well as the arbitrariness in the choice of time window. Nonetheless, different choices of time windows in our fitting consistently yield a value for $\beta$ that is smaller than 0.5. We thus conclude that spin transport at strong disorder is indeed subdiffusive.

As the disorder strength is decreased, the power $\beta$ increases continuously. For example, $\beta \approx 0.47$ for $g=0.2$ as shown in Fig.~\ref{fig:correlation}. Upon further decreasing the amount of disorder, $\beta$ appears to exceed 0.5 which implies superdiffusion. However, this conclusion is false. As the disorder strength decreases, the mean free path of the system $l_{\rm mfp}$ increases. At some point, the mean free path at that disorder strength becomes  the order of the system size used in our simulation $l_{\rm mfp} \sim L=20$, and thus simulations on small system sizes yield superdiffusive behaviors. Beyond that point, the system size accessible in our numerics is insufficient to draw any conclusion on the nature of transport in the thermodynamic limit and hence can no longer be trusted. On the other hand, at strong disorders $l_{\rm mfp}$ is typically much smaller than the system size, and hence simulations on moderate system sizes are good enough for inferring transport properties in the thermodynamics limit.

Although our numerics are inconclusive for weak disorders, the $g=1$ limit is well understood. At $g=1$, the system becomes two decoupled clean Heisenberg chains on the even and odd sites, respectively. This limit is integrable, and spin transport is superdiffusive with an exponent $\beta=\frac{2}{3}$~\cite{ljubotina2017spin, PhysRevLett.122.127202}. One may thus conjecture that upon adding disorder, transport becomes diffusive similar to the random field $XXZ$ chain with a MBL phase transition~\cite{PhysRevLett.117.040601}, although one needs a much larger system size to see diffusion.

\section{Summary and outlook}

In this work, we study themalization and spin transport in a disordered Floquet model with SU(2) symmetry. This model can be viewed as an extension of disordered Heisenberg Hamiltonians when energy conservation is removed, or an extension of Floquet-MBL models when an additional SU(2) symmetry is imposed. We find that, despite the absence of a true MBL phase, the model exhibits an extended nonergodic regime, which is characterized by both the spectral statistics and a direct comparison with ETH using expectation values of local observables. Moreover, we provide numerical evidence from the spin autocorrelation function indicating that spin diffusion at strong disorders is also anomalous.

Our result raises several interesting questions for future study. First, in Hamiltonian systems, a length scale beyond which resonances proliferate and the system eventually thermalizes can be extracted, using a real-space strong disorder renormalization group approach. While such a procedure is not directly applicable to Floquet systems where energy cannot be defined, is there a similar length scale controlling the ultimate thermalization in the strong disorder regime? Second, the eigenstates in strongly disordered Heisenberg chain can be well-approximated by tree tensor networks. What is the structure of the eigenstates in a Floquet system? The extended and yet nonergodic nature of the eigenstates also suggests multifractality that appears quite generically in many-body localized systems and Floquet systems with disorder~\cite{10.21468/SciPostPhys.4.5.025, PhysRevLett.123.180601}.
Finally, the transport properties of the strongly disordered Heisenberg chain has remained unexplored. It will be interesting to see if there is a subdiffusive regime there as well. We focus on $C_{zz}(t)$ in this work, but one can also look at other quantities such as the ac conductivity $\sigma(\omega)$, whose scaling exponent at low frequencies is in fact related to $\beta$.
Furthermore, it is desirable to identify the crossover from subdiffusion to diffusion by using different numerical methods that are amenable for much bigger system sizes, e.g. probing the steady-state current by coupling the system to leads~\cite{PhysRevLett.117.040601}.

\section*{Acknowledgments}

We thank Antonello Scardicchio for helpful discussions. Z.-C. Y. acknowledges funding by AFOSR FA9550-19-1-0399, ARO W911NF2010232. Z.-C. Y. is also supported by DoE BES Materials and Chemical Sciences Research for Quantum Information Science program (award No. DE-SC0019449), NSF PFCQC program, DoE ASCR Accelerated Research in Quantum Computing program (award No. DE-SC0020312), DoE ASCR Quantum Testbed Pathfinder program (award No. DE-SC0019040), AFOSR, ARO MURI, ARL CDQI, AFOSR MURI, and NSF PFC at JQI. M.C. acknowledges support from Alfred P. Sloan Foundation. Numerical calculations were performed on the Boston University Shared Computing Cluster, which is administered by Boston University Research Computing Services.


\bibliography{reference}



\end{document}